\def\tipo{2}

\def\av#1{\langle#1\rangle}
\newcommand{\omitit}[1]{}

\ifnum \tipo = 2
 \documentclass[aps,pre,twocolumn,showpacs]{revtex4}
 \def\figsize{8.5cm}
 \def\figsiz1{8cm}
\def \frontmatter{}

\else
 \documentclass[preprint,aps,pre,psfig]{revtex4}
 \def\figsize{12cm}
 \def\figsiz1{12cm}
 \def\frontmatter{}
 
\fi

\usepackage{graphicx}
\usepackage{dcolumn}
\usepackage{bm}
\usepackage{psfig} 

\begin{document}
\ifnum \tipo = 2
\else
  \draft
\fi  
\frontmatter
\title{Volatility of Linear and Nonlinear Time Series}
\author{Tomer~Kalisky$^1$, Yosef~Ashkenazy$^2$, and Shlomo Havlin$^1$}
\affiliation{
  $^1$ Minerva Center and Department of Physics, Bar-Ilan
  University, Ramat-Gan, Israel\\
  $^2$ Environmental Sciences, Weizmann Institute, Rehovot, Israel
}
\date{\today} 
\begin{abstract}
  { Previous studies indicate that nonlinear properties of Gaussian
    time series with long-range correlations, $u_i$, can be detected
    and quantified by studying the correlations in the magnitude
    series $|u_i|$, i.e., the ``volatility''. However, the origin for
    this empirical observation still remains unclear, and the exact
    relation between the correlations in $u_i$ and the correlations in
    $|u_i|$ is still unknown.  Here we find analytical relations
    between the scaling exponent of linear series $u_i$ and its
    magnitude series $|u_i|$.  Moreover, we find that nonlinear time
    series exhibit stronger (or the same) correlations in the
    magnitude time series compared to linear time series with the same
    two-point correlations.  Based on these results we propose a
    simple model that generates multifractal time series by explicitly
    inserting long range correlations in the magnitude series; the
    nonlinear multifractal time series is generated by multiplying a
    long-range correlated time series (that represents the magnitude
    series) with uncorrelated time series [that represents the sign
    series $sgn(u_i)$].  Our results of magnitude series correlations
    may help to identify linear and nonlinear processes in
    experimental records.  }
\end{abstract}


\ifnum \tipo = 2
\fi
\maketitle

\def\figureI{
\begin{figure}
\resizebox{\figsize}{!}{ \includegraphics{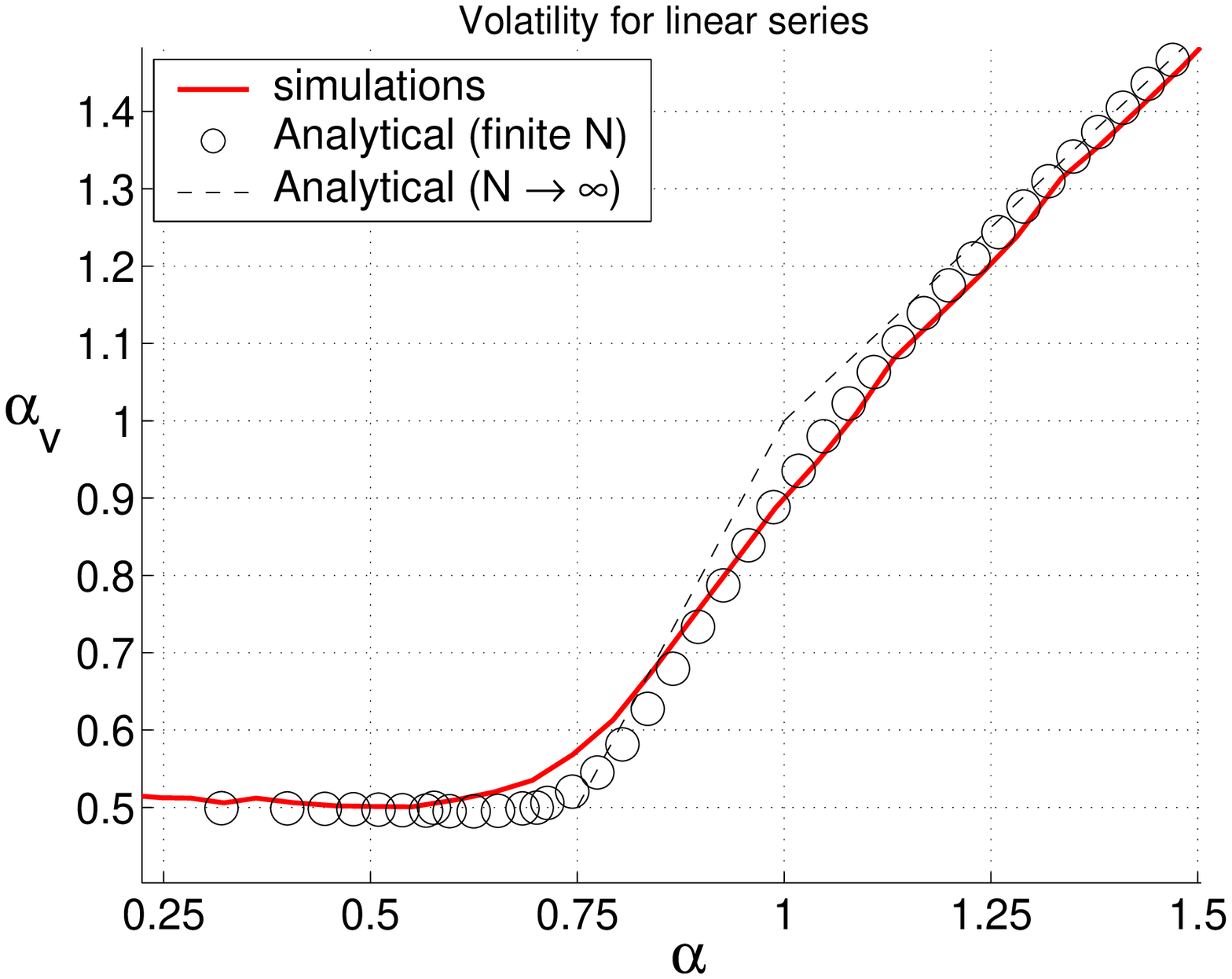} }              
\caption{\label{fig:vol_linear}
  Magnitude series scaling exponent $\alpha_v$ vs. the two-point
  correlation exponent $\alpha$ for linear sequences $u_i$. The solid
  line represents results for synthesized sequences of length
  $2^{15}$, averaged over 15 configurations, for $u_i^2$ and $|u_i|$
  (these two coincide). The circles represent the analytical
  reconstruction taking into account corrections due to finite size
  effects and non-stationarity.  Analytical results for $N\to \infty$
  are given by the dashed line.}
\end{figure}
}

\def\figureII{
\begin{figure}
\resizebox{\figsize}{!}{\includegraphics{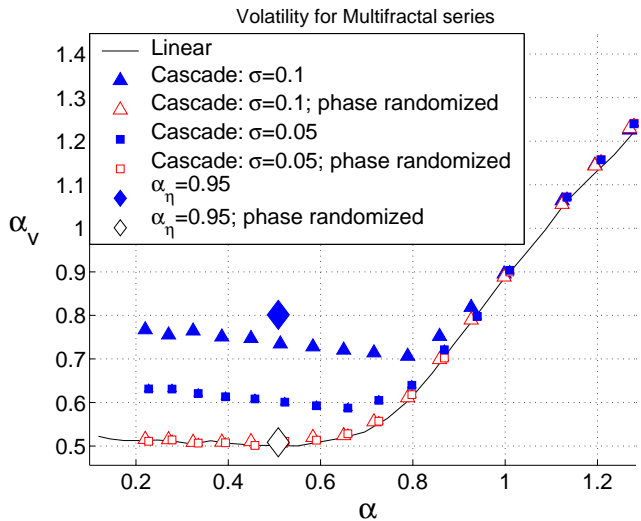}}
\caption{\label{fig:vol_multifractal} The magnitude series exponent
  $\alpha_v$ vs. the two-point correlation exponent $\alpha$ for multifractal
  series. The full triangles and squares represent sequences generated
  by the log-normal random cascade algorithm with $\sigma=0.1$ (triangles)
  and $\sigma=0.05$ (squares). The respective linear (phase randomized)
  surrogate data sequences are represented by empty symbols. The solid
  line indicates simulation results for linear sequences as derived in
  \cite{Ashkenazy-Havlin-Ivanov-et-al-2003:magnitude}, explained in
  section \ref{sec:linear}, and shown in
  Fig.~\protect\ref{fig:vol_linear}. The full diamond represents the
  scaling exponent of our multifractal model's sequences $u_i = \epsilon_i
  \eta_i$ with $\alpha_{\eta}=0.95$, while the empty diamond represents the
  respective scaling exponent of the surrogate data. All sequences are
  of length $2^{14}$ elements, and results were averaged over 15
  configurations. Error bars are smaller than symbol size.}
\end{figure}
}

\def\figureIII{
\begin{figure}
\vskip 3mm
\resizebox{\figsize}{!}{\includegraphics{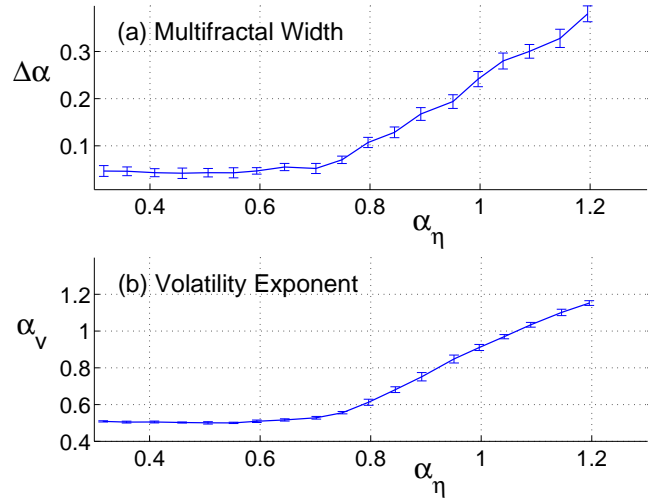}}
\caption{\label{fig:mrw_alpha_eta_vol_width_19} (a) Multifractal spectrum
  width and (b) volatility exponent $\alpha_v$ for sequences of the form
  $u_i = \epsilon_i \eta_i$ of length $2^{19}$, averaged over 15 configurations.
  The error bars indicate the mean $\pm$ 1 std. For $\alpha_{\eta}>0.75$ both
  volatility correlation exponent and the multifractal spectrum width
  of the series are increasing with $\alpha_{\eta}$. }
\end{figure}
}

\def\figureIV{
\begin{figure}
\vskip 5mm
\resizebox{\figsize}{!}{ \includegraphics{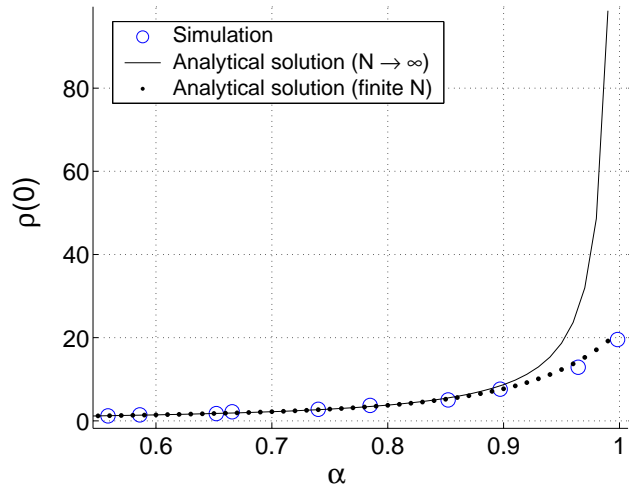} }              
\caption{\label{fig:ro_large}
  Correlation coefficient $\rho(0)$ (i.e. the variance $\av{u_i^2}$)
  for linear sequences of $N=50,000$ points. Circles indicate the
  simulation results. Dots represent analytical results for the
  variance calculated according to Eq. (\ref{equ:var_finite_N}), which
  takes into account the finite size effects. The solid line is the
  variance for $N \to \infty$. It can be seen that as $\alpha \to 1$
  the convergence becomes slower and finite size effects become more
  dominant [i.e., the convergence is non-uniform in the range $\alpha
  \in (0,1)$]. }
\end{figure}
}

\def\figureV{
\begin{figure}
\resizebox{\figsize}{!}{ \includegraphics{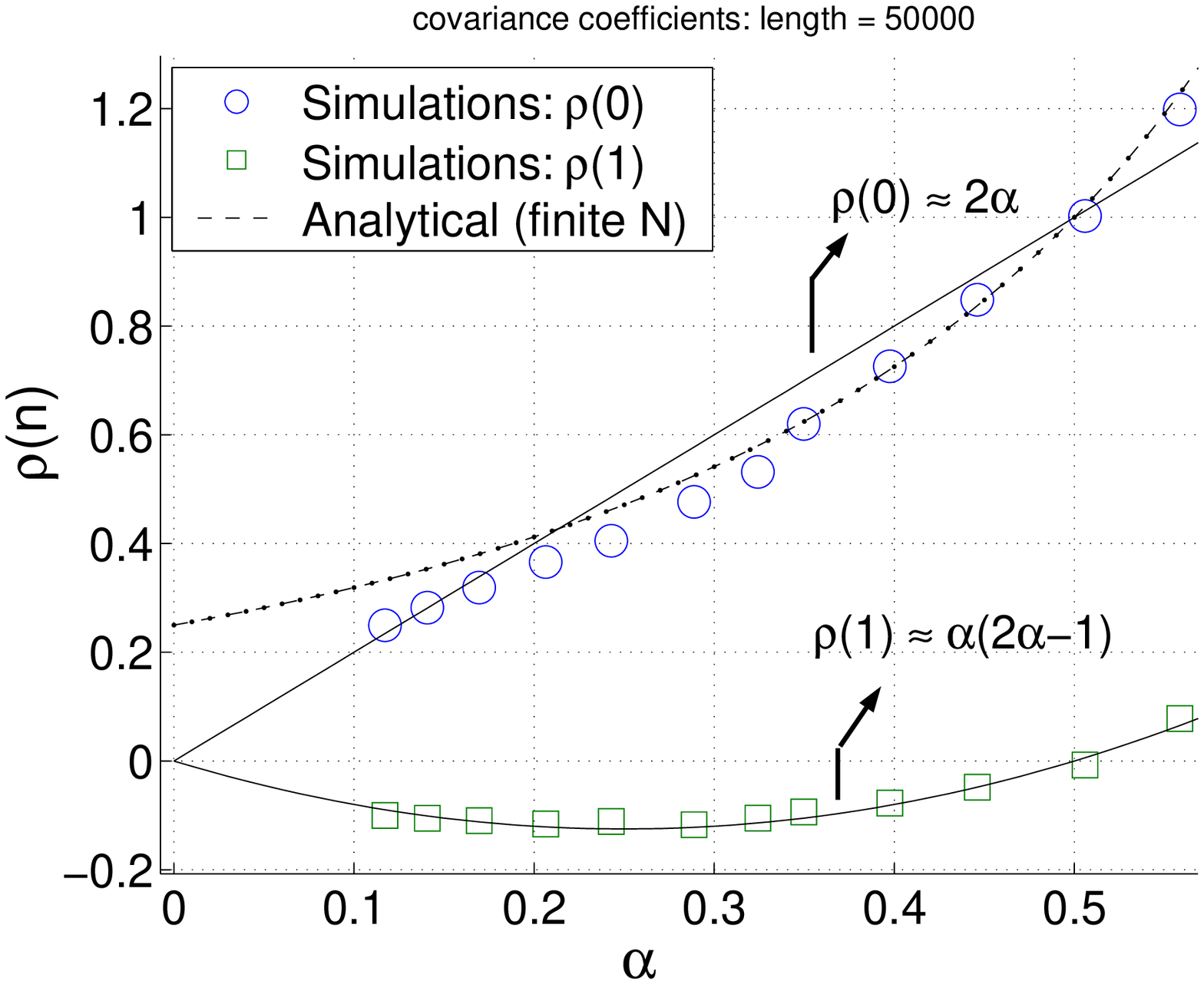} }              
\caption{\label{fig:ro_small}
  Correlation coefficients $\rho(0)$ (circles) and $\rho(1)$ (squares) for
  linear sequences of 50,000 points, in range $\alpha<1/2$.  The dashed
  line indicates the analytical results for $\rho(0)$, taking into
  account the finite series size effects, which approximately follows
  results for $N \to \infty$ (solid line). $\rho(1)$ is negative for $\alpha<1/2$
  indicating anti-correlations. The solid lines are the analytical
  expressions of
  Eq.~(\protect\ref{equ:correlation_definition_anti_corr}).}
\end{figure}
}

\section{\label{sec:Intro}Introduction}
Natural systems often exhibit irregular and complex behavior that at
first look erratic but in fact posses scale invariant structure
\cite[e.g.,][]{Shlesinger-1987:fractal,
  Peng-Havlin-Stanley-Goldberger-1995:quantification}.  In many cases this
nontrivial structure points to long-range temporal correlations, which
means that very far events are actually (statistically) correlated
with each other.  Long-range correlations are usually characterized by
scaling laws where the scaling exponents quantify the strength of
these correlations.  However, it is clear that the two-point
long-range correlations reveal just one aspect of the complexity of
the system under consideration and that higher order statistics is
needed to fully characterized the statistical properties of the system.

The two-point correlation function is usually used to quantify the
scale invariant structure of time series (long-range correlations),
while the $q$-point correlation function quantifies the higher order
correlations. In some cases the $q$-point correlation function is
trivially related to the two-point correlation function, and the
scaling exponents of different moments are linearly dependent on the
second moment scaling exponent. These kind of processes are termed
``linear'' and ``monofractal'', since just a single exponent that
determines the two-point correlations (and thus the linear
correlations) quantifies the entire spectrum of $q$ order scaling
exponents. In other cases, the relation between the $q$-point
correlation function has nontrivial relation to the two-point
correlation function, and a (nontrivial) spectrum of scaling exponents
is needed to quantify the statistical properties of the system; these
processes are called ``nonlinear'' and ``multifractal''. The
classification into linear and nonlinear processes is important for
the understandings of the underlying dynamics of natural time series
and for models development. Moreover, the nonlinear properties of
natural time series may have practical diagnosis
use~\cite[e.g.,][]{Ashkenazy-Ivanov-Havlin-et-al-2001:magnitude}.

Recently a simple measure for nonlinearity of time series was
suggested \cite{Ashkenazy-Ivanov-Havlin-et-al-2001:magnitude}. Given a
time series $u_i$, the correlations in the magnitude series
(volatility) $|u_i|$ may be related (in some cases) to the nonlinear
properties of the time series; basically, when the magnitude series is
correlated the time series $u_i$ is nonlinear. It was also shown that
the scaling exponent of the magnitude series may be related, in some
cases, to the multifractal spectrum width. However, these observations
are empirical and the reasons underly these observations still remain
unclear.

Here we develop an analytical relation between the scaling exponent of
the original time series $u_i$ and the scaling exponent of the
magnitude time series $|u_i|$ for linear series. We show that when the
original time series is nonlinear, the corresponding scaling exponent
of the magnitude series is larger (or in some cases equal) compared to
that of linear series and that the correlations in the magnitude
series increase as the nonlinearity of the original series increases.
These relations may help to identify nonlinear processes and quantify
the nonlinearity strength. Based on these results we suggest a generic
model for multifractality by multiplying random signs with long-range
correlated noise, and show that the multifractal spectrum width and
the volatility exponent increase as these correlations become
stronger.

The paper is organized as follows: in Section \ref{sec:background} we
present some background regarding non-linear processes and magnitude
(volatility) series correlations. In Section \ref{sec:linear} we
develop an analytical relation between the original time series
scaling exponent $\alpha$ and the magnitude series exponent $\alpha_v$; we
confirm the analytical relation using numerical simulation.  We then
study in Section \ref{sec:multifractal} the relation between
volatility correlations and the multifractal spectrum width of several
multifractal models, and introduce a simple model that generates
multifractal series by explicitly inserting long range correlations in
the magnitude series. A summary of the results is given in Section
\ref{sec:Summary}.

\section{\label{sec:background}Nonlinearity and volatility correlations}

\subsection{\label{subsec:Linear-Series}Two-point correlations}
%
The long range correlations of a time series $\{u_i\}$ $(i=0,1,2,\ldots,N)$
can be evaluated using the two-point correlation function
$\av{u_iu_j}$ ($\langle \cdot \rangle$ stands for expectation value); when $u_i$ is
long-range correlated and stationary the two-point correlation
function is $\av{u_iu_j} \sim |i-j|^{-\gamma}$ ($0<\gamma <1$)
\cite{Taqqu-Teverovsky-Willinger-1995:estimators,Bunde-Havlin-1996:fractals}.
It is possible to get good estimation of the scaling exponent using
various methods, such as the power spectrum, Fluctuation Analysis (FA)
\cite{Peng-Buldyrev-Havlin-et-al-1994:mosaic}, Detrended Fluctuation
Analysis (DFA)\cite{Peng-Buldyrev-Havlin-et-al-1994:mosaic,
  Peng-Havlin-Stanley-Goldberger-1995:quantification, Bunde-Havlin-Kantelhardt-et-al-2000:correlated}, wavelet
transform \cite{Muzy-Bacry-Arneodo-1994:multifractal}, and others; see
\cite{Taqqu-Teverovsky-Willinger-1995:estimators} for more details.
The different techniques characterize the linear two point
correlations in a time series with a scaling exponent which is related
to the scaling exponent $\gamma$.

In this study we use the FA method for the analytical derivations
since this method is relatively simple. In the FA method the sequence
is treated as steps of a random walk; then the variance of its
displacement, $X_t=\sum_{i=0}^{t} u_i$, is found by averaging over
different time windows of length $t$. The scaling exponent $\alpha$ of the
series (also referred to as the Hurst exponent $H$) can be measured
using the relation $Var(X_t) = \langle {X_t}^2 \rangle \sim t^{2\alpha}$ where $Var(\cdot)$ is
the variance; the scaling exponent $\alpha$ is related to the correlation
exponent $\gamma$ by $2-\gamma=2\alpha$.

\subsection{\label{subsec:non-Linear} High order correlations}
A more complete description of stochastic process $\{u_i\}$ with a zero
mean is given by its multivariate distribution: $P(u_0,u_1,u_2,...)$.
It is equivalent to the knowledge stored in the correlation functions
of different orders \cite{Stratonovich-1967:estimators,
  vanKampen-1981:stochastic}: $\av{u_i}$, $\av{u_iu_j}$, 
$\av{u_iu_ju_k}$, $\av{u_iu_ju_ku_l}$, etc. In many cases it is useful
to use the cummulants of different orders $C_q$ which are related to
the $q$ order correlation function by \cite{vanKampen-1981:stochastic}:
\begin{eqnarray}
  C_1 &=& \av{u_i} \\
  C_2 &=& \av{u_iu_j} \\
  C_3 &=& \av{u_iu_ju_k} \\
  C_4 &=& \av{u_iu_ju_ku_l}-\av{u_iu_j}\av{u_ku_l} - \nonumber\\
  &&\av{u_iu_k}\av{u_ju_l}-\av{u_iu_l}\av{u_ju_k}, \label{cummulants}
\end{eqnarray}
and so on. 

For a linear process (sometimes referred to as ``Gaussian'' process),
all cummulants above the second are equal to zero (Wicks theorem)
\cite{vanKampen-1981:stochastic}. Thus, in this case, the two-point
correlation fully describes the process \cite{Feder-1988:fractals,
  Frisch-1995:turbulence}, since all correlation functions (of
positive and even order) may be expressed as products of the two-point
correlation function $\av{u_iu_j}$.

Processes that are nonlinear (or ``multifractal'') have nonzero high
order cummulants. The nonlinearity of these processes may be detected
by measuring the multifractal spectrum
\cite{Feder-1988:fractals,Parisi-Frisch-1985:on} using advanced
techniques, such as the Wavelet Transform Modulus Maxima
\cite{Muzy-Bacry-Arneodo-1994:multifractal} or the Multifractal DFA
(MF-DFA)
\cite{Kantelhardt-Zschiegner-Koscielny-Bunde-et-al-2002:multifractal}.
In MF-DFA we calculate the $q$ order correlation function of the
profile $X_t=\sum_{i=0}^{t} u_i$ and the partition function is $Z_q(t)
\equiv \langle |X_t|^q \rangle $. For time series that obey scaling laws the partition
function is $Z_q(t) \sim t^{q\alpha (q)} $. Thus, the ``spectrum'' of scaling
exponents $\alpha (q)$ characterizes the correlation functions of different
orders. For a linear series, the exponents $\alpha (q)$ will all give a
single value $\alpha $ for all $q$
\cite{Kantelhardt-Zschiegner-Koscielny-Bunde-et-al-2002:multifractal}.

\subsection{\label{subsec:volatility} Volatility correlations}

A known example for the use of volatility correlations (defined below)
is econometric time series
\cite{Liu-Gopikrishnan-Cizeau-et-al-1999:statistical}.  Econometric
time series exhibit irregular behavior such that the changes
(logarithmic increments) in the time series have a white noise
spectrum (uncorrelated). Nonetheless, the {\em magnitudes} of the
changes exhibit long-range correlations that reflect the fact that
economic markets experience quiet periods with clusters of less
pronounced price fluctuations (up and down), followed by more volatile
periods with pronounced fluctuations (up and down).  This type of
correlation is referred to as ``volatility correlations''.

Given a time series $u_i$, the magnitude (volatility) series may be
defined as $|\Delta u_i|=|u_{i+1}-u_i|$.  The scaling exponent of the
magnitude series is the volatility scaling exponent $\alpha_v$.
Correlations in the magnitude series are observed to be closely
related to nonlinearity and multifractality
\cite{Arneodo-Bacry-Muzy-1998:random,
  Ashkenazy-Ivanov-Havlin-et-al-2001:magnitude,
  Ashkenazy-Havlin-Ivanov-et-al-2003:magnitude}

In this paper we refer to ``volatility'' with two small differences.
First, we consider the {\em square} of the series elements rather than
their absolute values. According to our abservations, this
transformation has negligible effect on the scaling exponent $\alpha_v$,
but it substantially simplifies the analytical treatment. Second, for
simplicity, we also consider the series itself rather than the
increment series. That is: the volatility series is defined as $u_i^2$
rather than $|\Delta u_i|$. Note that in most applications the absolute
values of the increment series are considered instead of the absolute
values of the series itself, since the original series is mostly
nonstationary (defined below); here we overcome this problem by first
considering stationary series.

\subsection{\label{subsec:stationary} Stationary and nonstationary
  time series}

Series with $0<\alpha<1$ are {\em stationary}, that is, their correlation
function depends only on the {\em difference} between points $i$ and
$j$, i.e., $\av{u_iu_j}=f(|i-j|)$. Their variance is a finite
constant, that does not increase with the sequence length.  Sequences
with $\alpha>1$ are {\em non stationary} and have a different form of
correlation function where the correlation function depends also on
the absolute indices $i$ and $j$,
$\av{u_iu_j}=i^{2\alpha-2}+j^{2\alpha-2}-{|i-j|}^{2\alpha-2}$; see
\cite{Taqqu-Teverovsky-Willinger-1995:estimators}. Scaling exponents
of nonstationary series (or series with polynomial trends) may be
calculated using methods that can eliminate constant or polynomial
trends from the data \cite{Muzy-Bacry-Arneodo-1994:multifractal,
  Peng-Buldyrev-Havlin-et-al-1994:mosaic, Bunde-Havlin-Kantelhardt-et-al-2000:correlated}.

\section{\label{sec:linear}Volatility correlations of linear time series}

We proceed to study the relation between the volatility correlation
exponent $\alpha_v$ and the original scaling exponent $\alpha$ for
linear processes, both numerically and analytically.

\subsection{\label{subsec:simulation} Simulations}

We generate artificial long-range correlated linear sequences $u_i$
with different values of $\alpha$ in the range $\alpha \in (0,1.5]$ as
follows~\cite{Makse-Havlin-Schwartz-Stanley-1996:method}: (i) generate
Gaussian white noise series, (ii) apply Fourier transform on that series,
(iii) multiply the power spectrum $S(f)$ by $1/f^\beta$ where $\beta
=2\alpha -1$ and $f >0$, and (iv) apply inverse Fourier transform. The
resultant series is long-range correlated with a scaling exponent
$\alpha$. We measure the volatility scaling exponent $\alpha_v$,
i.e. the scaling exponent of $u_i^2$ and $|u_i|$, versus the original
scaling exponent. The results are plotted in
Fig.~\ref{fig:vol_linear}.

\ifnum\tipo=2
  \figureI
\fi

The simulations indicate that the dependence of $\alpha_v$ on $\alpha$
for linear series may be divided into three regions: for $\alpha<3/4$
we obtain $\alpha_v\approx 1/2$, for $\alpha>1.25$ we obtain
$\alpha_v\approx
\alpha$, while for $0.75<\alpha<1.25$ there is a transition region
between those two behaviors. These results were obtained using the DFA
method which can handle nonstationary time series.

It is important to note that in Fig.~\ref{fig:vol_linear} we used the
absolute value of the series $u_i$, to calculate the volatility
scaling exponent. Nevertheless, the usual method for calculating the
volatility scaling exponent is performed by taking the absolute value
of the {\em differences} series, $\Delta u_i$, rather than $u_i$ itself
\cite{Ashkenazy-Ivanov-Havlin-et-al-2001:magnitude,
  Ashkenazy-Havlin-Ivanov-et-al-2003:magnitude}. The reason for this
is that in many cases the given time series has an exponent in the
range $\frac{1}{2}<\alpha<1\frac{1}{2}$. By differentiating the sequence we
get a new sequence $\Delta u_i$ with an exponent
$\tilde{\alpha}=\alpha-1<\frac{1}{2}$.  According to our analysis, if the
sequence is linear, the volatility exponent for this series will be
$\tilde{\alpha_v}=\frac{1}{2}$ (whereas for the original series $\alpha_v$ may
be higher than $\frac{1}{2}$ even for linear data). Thus, if the
scaling exponent of the magnitude of the {\em differences} series, $|\Delta
u_i|$, differs from 1/2 this is an indication for nonlinearity.

We note that for $\alpha \gg 1$ the series is highly non-stationary,
i.e., it is most of the time either above or below $0$, apart from a
discrete set of crossing points. The behavior of the series $u_i$ is
not very different than the behavior of its absolute value $|u_i|$,
and therefore it is not surprising that $\alpha_v=\alpha$.

\subsection{\label{subsec:analytical}Analytical Treatment}

Let us consider a Gaussian distributed linear sequence $u_i$ of length
$t$ with scaling exponent $\alpha$. For simplicity, we assume that the
sequence is stationary. Consider the magnitude series: $v_i = u_i^2$.
In order to calculate the magnitude series scaling exponent $\alpha_v$ we
will calculate the variance of the displacement $V_t=\sum_{i=0}^{t} v_i$:
\begin{eqnarray}
\begin{array}{c}
Var(V_t) = \av{V_t^2} - \av{V_t} \av{V_t} = \nonumber\\\\
= \sum_{i=0}^{t} \sum_{j=0}^{t} [\av{v_i v_j} - \av{v_i} \av{v_j}] =
\nonumber\\\\
= \sum_{i=0}^{t} \sum_{j=0}^{t} [\av{u_i^2 u_j^2} - \av{u_i^2}
\av{u_j^2}].
\end{array}
\end{eqnarray}
Because the series $u_i$ is linear, the fourth cumulant is $C_4=0$
(Wicks theorem), and by using Eq.~(\ref{cummulants}) we get,
\begin{equation}
\av{u_i^2 u_j^2} = 
\av{u_i^2} \av{u_j^2} + 2{\av{u_iu_j}^2},
\end{equation}
and thus,
\begin{eqnarray}
\label{equ:sumsum_sqr}
Var(V_t) = 
2 \sum_{i=0}^{t} \sum_{j=0}^{t} {\av{u_iu_j}}^2. \nonumber
\end{eqnarray}
Substituting the two-point correlation function for long-range correlated data:
\begin{eqnarray}
\label{equ:correlation_definition}
\rho(i-j) = \av{u_i u_j} \sim \left\{
\begin{array}{ll}
|i-j|^{-\gamma}     &     \;i \not= j  \\
1         &     \;i=j        \\
\end{array}
\right.
\end{eqnarray}
we obtain 
\begin{eqnarray}
Var(V_t) \sim t + \sum_{i \not= j}^{t} |i-j|^{-2\gamma} \sim \nonumber\\
\sim  t + t^{-2\gamma+2}.
\end{eqnarray}
Since $2-\gamma=2\alpha$ the above expression becomes 
\begin{eqnarray}
\label{equ:final_linear_terms}
\av{V_t^2} \sim t + t^{4\alpha-2} = t^{2\alpha_v}.
\end{eqnarray}
For $\alpha<\frac{3}{4}$ the first term, $t$, is dominant and for $t
\to \infty $ we obtain $\alpha_v=\frac{1}{2}$. Otherwise the second
term, $t^{4\alpha-2}$, is dominating and thus $\alpha_v \approx
2\alpha-1$. 
However, the simulation results (Fig.~\ref{fig:vol_linear}) deviate
from $\alpha_v \approx 2\alpha-1$ as $\alpha \to 1$. This is because
as $\alpha \to 1$, logarithmic and polynomial corrections due to
strong finite size effects and non-stationarity must be taken into
account in our calculations (i.e., the variance of the sequence
depends on its length; see Appendix~\ref{app:finite-size}).  This is
done by dividing $\av{V_t^2}$ [Eq.~(\ref{equ:final_linear_terms})] by
the variance of the sequence:
\begin{eqnarray}
  \label{variance}
Var(u_i) = \frac{1}{1-\alpha}2^{2\alpha-2}(1-t^{2\alpha-2}) \sim 
\left\{
\begin{array}{ll}
{\rm const}              &     \;\alpha \ll 1  \\
\ln t               &     \;\alpha = 1    \\
t^{2\alpha-2}       &     \;\alpha \gg 1  \\
\end{array}
\right.
\end{eqnarray}

This modification yields an $\alpha_v$ that is very close to the one
obtained from the numerical simulation (in the transition region
$0.75<\alpha<1.25$, and also for $\alpha>1.25$ with $\alpha_v=\alpha$;
see Fig.~\ref{fig:vol_linear}).  
The relation $\alpha_v=\alpha$ for $\alpha>1.25$ can now be proved
analytically: It is noticeable that the dominant scaling term of
$\av{V_t^2}$ for the nonstationary case is proportional to
$t^{4\alpha-2}$ [Eq.~(\ref{equ:final_linear_terms})].  Dividing by the
variance term, $t^{2\alpha-2}$ [Eq.~(\ref{variance})], yields
$t^{2\alpha}\sim t^{2\alpha_v}$ and hence $\alpha=\alpha_v$.

\section{\label{sec:multifractal}Volatility correlations and the
  Multifractal spectrum width}

\subsection{\label{subsec:cascade}Random multifractal cascades}

%
Following \cite{Ashkenazy-Ivanov-Havlin-et-al-2001:magnitude,
  Ashkenazy-Havlin-Ivanov-et-al-2003:magnitude}, we study the relation
  between the volatility scaling exponent $\alpha_v$ and the
  multifractal spectrum width of nonlinear multifractal time
  series. We generate artificial noise with multifractal properties
  according to the algorithm proposed in
  \cite{Arneodo-Bacry-Muzy-1998:random}. The algorithm is based on
  random cascades on wavelet dyadic trees. The multifractal series is
  constructed by building its wavelet coefficients at different scales
  recursively, where at each stage the coefficients of the coarser
  scale are multiplied by a random variable $W$ in order to built the
  coefficients of the finer scale. Note that we now consider the
  increments series, hence the generated time series is
  stationary. The multifractal spectrum $f(\alpha)$ depends on the
  statistical properties of the random variable $W$.

We choose $W$ to follow log-normal distribution, such that the
$\ln{|W|}$ is normally distributed with $\mu$ and $\sigma^2$ being the mean
and variance. For this case the multifractal spectrum $f(\alpha)$ is known
analytically \cite{Arneodo-Bacry-Muzy-1998:random} and by assigning
$f(\alpha)=0$ it possible to obtain $\alpha_{min,max}$,
\begin{eqnarray}
\alpha_{min}&=-&\frac{\sqrt{2} \, \sigma}{\sqrt{\ln{2}}} - \frac{\mu}{\ln{2}}\\
\alpha_{max}&=&\frac{\sqrt{2} \, \sigma}{\sqrt{\ln{2}}} - \frac{\mu}{\ln{2}}.
\end{eqnarray}
Thus, the multifractal width, $\Delta \alpha =\alpha_{max}-\alpha_{min}=2\frac{\sqrt{2} \,
  \sigma}{\sqrt{\ln{2}}}$, depends just on $\sigma$ while the scaling exponent
$\alpha(0)$ depends on $\mu$, i.e. $\alpha(0)=-\frac{\mu}{\ln 2}$
\cite{Arneodo-Bacry-Muzy-1998:random}.

Using the above algorithm, we generate multifractal time series with a
fixed multifractal width $\Delta \alpha$ (by fixing $\sigma$) and
different scaling exponents $\alpha(q=2)$ (by changing $\mu$), and
calculate their volatility exponents $\alpha_v$ (see
Fig. \ref{fig:vol_multifractal}) \cite{remark1}. We perform the same
analysis for their respective surrogate time series, which are
linearized series (after phase randomization) that have the same
two-point correlations with exponent $\alpha(q=2)$ as the original
series \cite{Schreiber-Schmitz-2000:surrogate}; note that $\alpha (q)$
is the $q$-point correlation exponent. We find that the volatility
exponent $\alpha_v$, calculated in section \ref{sec:linear} for the
linear case, is the lower bound for all multifractal sequences
(studied here) with same $\alpha(q=2)$.  Nonetheless, for $\alpha(q=2)
> 1$, i.e., for nonstationary series, $\alpha_v=\alpha(q=2)$ as in
linear series. It is clearly seen in Fig. \ref{fig:vol_multifractal}
that for stationary time series ($\alpha(2)<1$), the volatility
correlations increase as the multifractal spectrum width becomes
wider, or alternatively, as the nonlinearity of the original series
strengthens.

\ifnum\tipo=2
  \figureII
\fi

\subsection{\label{subsec:mrw}A simple model for multifractality}
We now propose a simple model for generating multifractal records,
based on the property that multifractal series exhibit long range
correlations in the volatility series.
Following \cite{Bacry-Delour-Muzy-2001:multifractal}, we multiply a
long range correlated series $\eta_i$ (with a scaling exponent
$\alpha_{\eta}>0.75$) with a series of uncorrelated random signs
$\epsilon_i= \pm 1$.  The resultant series, $u_i = \epsilon_i \eta_i$,
has a two-point correlation exponent $\alpha=1/2$ because of the
random signs $\epsilon_i$. The magnitude exponent $\alpha_v$ is the
same as the magnitude exponent $\alpha_{v,\eta}$ for $\eta_i$, because
$|u_i|=|\eta_i|$. Thus, using our results from
Section~\ref{sec:linear}, if we take $\alpha_{\eta}>0.75$ we get a
sequence with $\alpha=1/2$ and
$\alpha_v \approx 2\alpha_{\eta}-1 > 1/2$
(see Fig.~\ref{fig:vol_multifractal}, full diamond symbol for
$\alpha_{\eta}=0.95$). Note that in Fig.~~\ref{fig:vol_multifractal}
the theoretical value of $\alpha_v=2\times 0.95-1=0.9$ is higher than
that of the numerical estimation $\alpha_v=0.8$, most probably due to
finite size effects.

According to our derivation in Section~\ref{sec:linear}, this sequence
is nonlinear/ multifractal (because linear series with a two-point
correlation exponent $\alpha=1/2$ should have $\alpha_v=1/2$).
Indeed, one can see from Fig.~\ref{fig:mrw_alpha_eta_vol_width_19}
that the multifractal width for this model increases as
$\alpha_{\eta}$ increases beyond 0.75.

Natural processes are often characterized by complex nonlinear and
multifractal properties. However, the underlying mechanisms of these
processes are usually not so well understood. Several prototypes for
multifractal processes include, e.g., (i) the energy cascade model
describing turbulence \cite{Frisch-1995:turbulence,
  Muzy-Bacry-Arneodo-1994:multifractal}, (ii) the {\em universal
  multifractal process} usually used to generally explain geophysical
phenomena \cite{Lovejoy-Schertzer-1986:scale,
  Schmitt-Lovejoy-Schertzer-1995:multifractal}, and (iii) the
turbulence-like model for heart rate variability
\cite{Lin-Hughsom-2001:modeling}.

The multifractal model described in this section is a simple model
with known properties that may help to gain better understanding
of\omitit{ gaining basic understandings regarding} multifractal
processes. The model consists of two components as follows, (i) a
random series (which can be also any other long-range correlated
series) that may represent fast processes of a natural system, which
as a first approximation may be regarded as a white noise, interacting
with (ii) a long-range correlated process that may represents a slow
modulation of the natural system. This interaction results in episodes
with less volatile fluctuations followed by episodes with more
volatile fluctuations. In the context of heart-rate variability, the
fast component may represents the parasympathetic branch of the
autonomic nervous system while the slow process may represents the
sympathetic and the hormonal activities. In the context of geophysical
phenomena, the fast component may represents the fast atmospheric
processes
\cite{Hasselmann-1976:stochastic} while the slow process may represent
the relatively slow oceanic processes.
Our model
can also be used to model other complex systems like economy and
heartbeat dynamics.

\ifnum\tipo=2
  \figureIII
\fi

\section{\label{sec:Summary} Summary}
We study the behavior of the magnitude series scaling exponent $\alpha_v$
versus the original two-point scaling exponent $\alpha$ for linear and
nonlinear (multifractal) series. We find analytically and by
simulations that for linear series the dependence of $\alpha_v$ on $\alpha$ may
be divided into three regions: for $\alpha<3/4$ the volatility exponent is
$\alpha_v=1/2$, for $\alpha>1.25$ the volatility exponent is $\alpha_v=\alpha$, while for
$0.75<\alpha<1.25$ there is a transition region in which logarithmic
corrections due to finite size effects and non-stationarity are
dominant.

The results presented here provide the theory for the relation found
previously \cite{Ashkenazy-Ivanov-Havlin-et-al-2001:magnitude,
Ashkenazy-Havlin-Ivanov-et-al-2003:magnitude} between multifractality
and the scaling exponent of the magnitude of the differences series
(volatility). This relation provides a simple method for preliminary
detection and quantification of nonlinear time series, a procedure
which usually requires relatively complex techniques.

We also study the volatility of some known models of multifractal time
series, and find that their magnitude scaling exponent is bounded from
below by $\alpha_v$ of the corresponding phase randomized linear
surrogate series.

Based on the above findings, we propose a simple model that generates
multifractality by explicitly inserting long range correlations
($\alpha_{\eta}>0.75$) into the magnitude series. This model may serve
as a generic model for multifractality and may help to gain
preliminary understandings of natural complex phenomena. The model,
which involves interaction between fast and slow components, may
represents natural fast processes that interact with slower
processes. In addition, the simplicity of the model may help to
identify these processes more easily in experimental records.


\section*{Acknowledgments}
We wish to thank the Israeli Center for Complexity Science for
financial support, and Yshai Avishai for useful discussions.

\appendix
\section{\label{app:finite-size}Finite Size Effects and Non-Stationarity near $\alpha=1$}
 
A linear time sequence with scaling exponent $\alpha$ can be generated
by filtering Gaussian white noise such that the power spectrum will
be \cite{Makse-Havlin-Schwartz-Stanley-1996:method}:
\begin{eqnarray}
S(f) \sim
\left\{
\begin{array}{ll}
0                   &     \;f=0          \\
\frac{1}{|f|^\beta}   &     \;f \neq 0     \\
\end{array}
\right.
\end{eqnarray}
where $\beta=2\alpha-1$.
Assume a signal $u_i$ of $N$ discrete points sampled at time intervals
$\Delta t$. The power spectrum consists of $N$ points in the frequency
range $(-\frac{1}{2 \Delta t},\frac{1}{2 \Delta t}]$ with intervals of $\Delta f =
\frac{1}{N \Delta t}$. Thus, looking only at the positive frequencies, the
minimal frequency (without loss of generality) is $\frac{\Delta f}{2} =
\frac{1}{2N \Delta t}$. The variance of the signal is the total area under
the power spectrum:
\begin{eqnarray}
Var(u_i) = 
2 \int_{\frac{1}{2N \Delta t}}^{\frac{1}{2 \Delta t}} S(f)df = 
2 \int_{\frac{1}{2N \Delta t}}^{\frac{1}{2 \Delta t}}
\frac{1}{f^{2\alpha-1}} df
\end{eqnarray}
Assuming $\Delta t = 1$, for $\alpha=1$ the variance is,
\begin{eqnarray}
Var(u_i) = 2 \ln N.
\end{eqnarray}
Thus, the variance diverges logarithmically for $\alpha=1$.

For $\alpha \neq 1$ the variance is
\begin{eqnarray}
\label{equ:var_finite_N}
Var(u_i) = \frac{1}{1-\alpha}2^{2\alpha-2}(1-N^{2\alpha-2}).
\end{eqnarray}
From (\ref{equ:var_finite_N}) follows: For $\alpha<1$ the variance
converges, and for $\alpha>1$ it diverges.
\omitit{Thus, the variance converges for $\alpha<1$ and diverges for $\alpha>1$.}

\textbf{Non-stationarity:} For $\alpha \geq 1$ the variance {\em diverges}
with the sequence length $N$, because of the singularity in the power
spectrum, and the sequence is non-stationary. For $\alpha>1$ the divergence
is power-law, i.e. $Var(u_i) \sim N^{2\alpha-2}$, while at $\alpha=1$ the
divergence is logarithmic.

\textbf{Finite size effects:} For $\alpha < 1$ the variance converges to
a finite constant so the sequence is stationary, but as $\alpha
\to 1$ this convergence becomes slower.
This means that as $\alpha \to 1$, larger and larger sequence lengths
$N$ are required so that the variance will indeed converge to a
constant value (see Fig.~\ref{fig:ro_large}). This argument also holds
for other values of the correlation functions $\rho(n)$,
$n=0,1,...,\infty$, although in a more moderate way.

\ifnum\tipo=2
  \figureIV
\fi
  
The strong finite size effects around $0.75<\alpha<1.25$ and the
non-stationarity at $\alpha \geq 1$ have to be taken into account when
calculating the magnitude series scaling exponent $\alpha_v$. This is done
by dividing the volatility fluctuation function $\av{V_t^2}$ by the
variance of the sequence given in Eq.~(\ref{equ:var_finite_N})
\cite{remark2}.

For $N \to \infty$ the finite size effects disappear and $\alpha_v$ converges to
its theoretical value (see Fig.~\ref{fig:vol_linear}). This
convergence is extremely slow and becomes weaker as we approach $\alpha \approx
1$.

%
For completeness, we show in Fig.~\ref{fig:ro_small} the correlation
coefficients $\rho(n=i-j)$ for $\alpha<1/2$. In this regime the sequences
exhibits short range anti-correlations as can be seen in
Fig.~\ref{fig:ro_small}. The expression of the correlation function
for $\alpha<1/2$ is approximately
\cite{Taqqu-Teverovsky-Willinger-1995:estimators}
\begin{eqnarray}
\label{equ:correlation_definition_anti_corr}
\rho(i-j) = \av{u_i u_j} \sim \left\{
\begin{array}{ll}
\alpha(2\alpha-1)|i-j|^{2\alpha-2}     &     \;i \not= j  \\
2\alpha                                &     \;i=j.        \\
\end{array}
\right.
\end{eqnarray}

\ifnum\tipo=2
  \figureV
\fi



\begin{thebibliography}{26}
\bibitem{remark1} Note that in the DFA notation the expressions for
  $\alpha$ should be larger by one than those of
  \protect\cite{Arneodo-Bacry-Muzy-1998:random}; however, since here
  we consider the increments, the series exponent is reduced back by
  one, thus compensating the DFA integration.
\bibitem{remark2} FA and DFA actually measure the scaling of the
  fluctuations for window sizes ranging from 1 to t. Fluctuations for
  windows of size t are given by $\av{X_t^2}$, while fluctuations for
  windows of size 1 are actually the variance of the sequence. Thus,
  the scaling exponent is approximated by $[\ln \av{X_t^2}^{1/2}-\ln
  \av{X_1^2}^{1/2}]/\ln t=[\ln \av{X_t^2}^{1/2}-\ln
  var(u_i)^{1/2}]/\ln t=\ln [\av{X_t}/var(u_i)]/2\ln t$.  Therefore,
  in cases where the variance is not constant, the fluctuation
  function $\av{V_t^2}$ should be normalized by the variance.
\end{thebibliography}

\ifnum\tipo=1
  \figureI
  \figureII
  \figureIII
  \figureIV
  \figureV
\fi

\end{document}